\documentclass[a4paper]{jpconf}

\usepackage{dcolumn}
\usepackage{bm}
\usepackage{amsmath, amssymb}
\usepackage{xspace}
\usepackage{comment}
\usepackage{bbold}
\usepackage{braket}

\ifx\pdfoutput\undefined
\usepackage[dvipdfmx]{graphicx}
\usepackage[dvipdfmx]{hyperref}
\usepackage[dvipdfmx]{color}
\usepackage[dvipdfmx]{xcolor}
\else
\usepackage{graphicx}
\usepackage{hyperref}
\usepackage{color}
\usepackage{xcolor}
\fi

\newcommand{\QQ}{\mathcal{Q}}
\newcommand{\vp}{\varphi}
\newcommand{\tvp}{\tilde{\varphi}}

\begin{document}
\title{Phase shift, ellipticity, angle, and topological number in skyrmion lattices}

\author{Kotaro Shimizu$^1$, Shun Okumura$^2$, Yasuyuki Kato$^1$, and Yukitoshi Motome$^1$}

\address{$^1$ Department of Applied Physics, The University of Tokyo, Tokyo 113-8656, Japan}
\address{$^2$ Institute for Solid State Physics, University of Tokyo, Kashiwa 277-8581, Japan}

\ead{k.shimizu@aion.t.u-tokyo.ac.jp}

\begin{abstract}
We theoretically study skyrmion lattices realized in a Kondo lattice model on a triangular lattice, 
focusing on the phase, ellipticity, and angle of the constituent multiple-$Q$ waves. 
Analyzing the numerical data obtained in the previous study [Ozawa R, Hayami S and Motome Y 2017 {\it Phys. Rev. Lett.} {\bf 118} 147205], we extract these parameters for the two types of skyrmion lattices with the skyrmion number of $1$ and $2$. 
We show that the topological transition between the two skyrmion lattices driven by an external magnetic field is accompanied by significant modulations of all three parameters.
\end{abstract}

\section{Introduction \label{sec:1}}

The magnetic skyrmion lattice (SkL) is a two-dimensional topological spin texture~\cite{Bogdanov1989,Bogdanov1994, Roessler2006, Muhlbauer2009, Yu2010}.  
It is described by a superposition of multiple spin density waves, and hence, called a multiple-$Q$ spin state. 
In such a state, the topological property can be changed by modulating the constituent waves. 
For instance, the symmetry of the spin textures and the skyrmion number $N_{\rm sk}$, which characterizes the topology of the SkL, are changed by phase shift in the constituent waves~\cite{Kurumaji2019,Hayami2020phase,Hayami2021locking}. 
Since there are many other parameters in the superpositions, further studies are desired to establish the way of controlling the topological property of multiple-$Q$ states. 

In this report, we clarify the evolution of the superposed waves in the SkLs discovered in the previous study~\cite{Ozawa2017}. 
Analyzing the numerical data of the real-space spin configurations, we extract the phases of the constituent waves, 
and the ellipticity and the angle of the propagating planes for two types of SkLs with $|N_{\rm sk}|=1$ and $2$. 
We show that all these three parameters change abruptly at the topological phase transition from $|N_{\rm sk}|=2$ to $1$ while increasing the magnetic field, indicating that the constituent waves are modulated through the topological change.

\section{Model and method \label{sec:2}}

We analyze the spin configurations numerically obtained in the ground state of the Kondo lattice model on a triangular lattice~\cite{Ozawa2017}.
The Hamiltonian is given by
\begin{eqnarray}
\mathcal{H}&=&-\sum_{{\bf r},{\bf r}',\sigma}t_{{\bf r} {\bf r}'}
\hat{c}^{\dag}_{{\bf r}\sigma}\hat{c}_{{\bf r}'\sigma}
-J\sum_{{\bf r},\sigma,\sigma'} {\bf S}_{{\bf r}}\cdot \hat{c}^{\dag}_{{\bf r} \sigma} 
\boldsymbol{\sigma}_{\sigma\sigma'} \hat{c}_{{\bf r} \sigma'} 
-h\sum_{{\bf r}} S_{{\bf r}}^z, 
\label{eq:KLmodel}
\end{eqnarray}
where the operator $\hat{c}^{\dag}_{{\bf r} \sigma}$ $(\hat{c}_{{\bf r} \sigma})$ creates (annihilates) an electron with spin index 
$\sigma=\pm$ at site ${\bf r}$ and $\boldsymbol{\sigma}$ is the vector of Pauli matrices; 
the localized spins ${\bf S}_{{\bf r}}$ are treated as classical vectors with $|{\bf S}_{{\bf r}}|=1$. 
See \cite{Ozawa2017} for the details of the parameters. 

In the previous study, two types of SkLs were found in the model (\ref{eq:KLmodel}): SkL with an unusual skyrmion number of $N_{\rm sk}=2$ in the low field region including at zero field, and SkL with $N_{\rm sk}=1$ in an intermediate field region. 
Since both of them are regarded as superpositions of three spin density waves (i.e., $3Q$ states) characterized by the wave vectors 
${\bf q}_{\eta}=R_{\triangle}^{\eta-1}(\pi/3,0)^{\mathsf{T}}$ ($\eta=1,2,3$), 
where $R_{\triangle}$ is a two-dimensional $\frac{2\pi}{3}$-rotation operator and $\mathsf{T}$ denotes the transpose of the vector, we assume the following functional form to approximately describe them: 
\begin{align}
&&{\bf S}^{3Q}_{{\bf r}}(\{\vp_\eta\}, m, \theta, \epsilon, \Gamma; \hat{\bf n}, \xi)
\propto R\left(\hat{\bf n}, \xi\right) \left[
\sum_{\eta=1}^3~\frac{1}{\sqrt{3}}\left( {\bf e}_{\eta}^1(\theta, \Gamma)\cos\QQ_{\eta} 
+ \epsilon{\bf e}_{\eta}^2(\theta, \Gamma)\sin\QQ_{\eta} \right) + m\hat{\bf z} \right], 
\label{eq:3q_fit_ansatz}
\end{align}
where 
$R(\hat{\bf n},\xi)$ denotes the three-dimensional rotation matrix about the unit vector $\hat{\bf n}$ with the angle $\xi$, 
${\bf e}_{\eta}^1(\theta, \Gamma)=\left[R\left(\hat{\bf z}, (-1)^{\Gamma}\frac{2\pi}{3}\right)\right]^{\eta-1}(0, (-1)^{\Gamma}\sin\theta, \cos\theta)^{\mathsf{T}}$, 
${\bf e}_{\eta}^2(\theta, \Gamma)=\left[R\left(\hat{\bf z}, (-1)^{\Gamma}\frac{2\pi}{3}\right)\right]^{\eta-1}(0, -(-1)^{\Gamma}\cos\theta, \sin\theta)^{\mathsf{T}}$, 
$\QQ_{\eta}={\bf q}_{\eta}\cdot{\bf r}+\varphi_{\eta}$, $\Gamma$ takes $0$ or $1$, 
and $\hat{\bf z}=(0,0,1)^{\mathsf{T}}$. 
In (\ref{eq:3q_fit_ansatz}), $\vp_{\eta}$, $\epsilon$, and $\theta$ denote the phase, the ellipticity, and the angle between the $z$ axis and the major axis of the ellipse, respectively, which are relevant parameters to the energy of (\ref{eq:3q_fit_ansatz}) for the model (\ref{eq:KLmodel}).
We note that the model in (\ref{eq:KLmodel}) does not change the energy by flipping all the $y$ components of 
the localized spins, which corresponds to the change of $\Gamma$ between $0$ and $1$ in (\ref{eq:3q_fit_ansatz}). 
This operation reverses the sign of $N_{\rm sk}$. 

For the spin configurations obtained numerically, $\{{\bf S}^{\rm GS}_{{\bf r}}\}$, we optimize the parameters in (\ref{eq:3q_fit_ansatz}) to minimize the cost function defined by
\begin{eqnarray}
U(\{\vp_\eta\}, m, \theta, \epsilon, \Gamma; \hat{\bf n}, \xi)=\frac{1}{N}\sum_{{\bf r}} 
\left(1-{\bf S}^{\rm GS}_{{\bf r}} \cdot {\bf S}^{3Q}_{{\bf r}}(\{\vp_\eta\}, m, \theta, \epsilon, \Gamma; \hat{\bf n}, \xi) \right),  
\label{eq:3q_cost} 
\end{eqnarray}
where $N$ is the number of sites. 
In the following, we focus on the optimal values of the phases $\varphi_{\eta}^{*}$, the ellipticity $\epsilon^*$, and the angle $\theta^*$. 
For the phases, paying attention to the sixfold rotational symmetry of the model, the symmetry by the transformation from $\vp_{\eta}$ to $-\vp_{\eta}$, and the $2\pi$ periodicity of $\sum_{\eta}\vp_{\eta}^{*}$, we define the sum of $\vp_{\eta}^{*}$ in the form of 
\begin{eqnarray}
\tvp = \pi - \left|\pi - {\rm Mod}\left[\sum_{\eta=1}^{3} \vp^{*}_{\eta}, 2\pi\right]\right|. 
\label{eq:phase_sum}
\end{eqnarray}

\section{Results \label{sec:3}}

\begin{figure}[tb]
\centering
\includegraphics[width=\columnwidth]{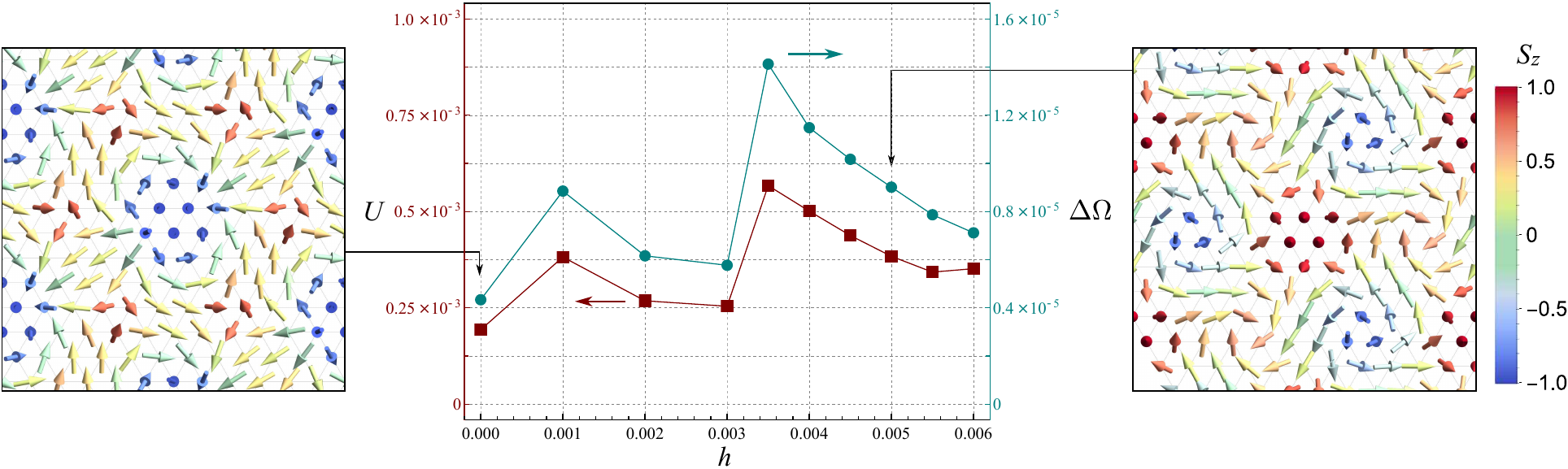}
\caption{\label{fig:1}
Magnetic field dependences of $U$ in (\ref{eq:3q_cost}) and the difference of the grand potentials between the optimal state and the numerically-obtained ground state, $\Delta\Omega=\Omega^{3Q}-\Omega^{\rm GS}$. 
The real-space spin configurations obtained by the optimization, $\{{\bf S}_{{\bf r}}^{3Q}\}^*$, for $h=0.000$ and $0.005$ are plotted 
in the left and right panels, respectively; the arrows and color denote the $xy$ and $z$ components of $\{ {\bf S}_{{\bf r}}^{3Q} \}^*$, respectively. 
}
\end{figure}

Figure~\ref{fig:1} shows $h$ dependences of the cost function in (\ref{eq:3q_cost}) and the difference of the grand potentials between the optimal state $\{{\bf S}_{{\bf r}}^{3Q}\}^*$ and the numerically obtained ground state $\{{\bf S}_{{\bf r}}^{\rm GS}\}$, $\Delta\Omega=\Omega^{3Q}-\Omega^{\rm GS}$. 
Here, the grand potential is defined by $\Omega=E-\mu n$, where $E=\braket{\mathcal{H}}/N$ and $n=\sum_{{\bf r},  \sigma}\braket{c^{\dagger}_{{\bf r}\sigma}c_{{\bf r}\sigma}}/N$. 
In the entire region of $h$ calculated, we obtain sufficiently small values of $U < 0.006$, indicating that $\{{\bf S}_{{\bf r}}^{3Q}\}^*$ gives a good approximation to $\{{\bf S}^{\rm GS}_{{\bf r}}\}$, as exemplified in the left and right panels for $h=0.000$ and $0.005$, respectively. 
$\Delta\Omega$ is also small enough, whose relative error is less than $0.5 \%$ for all $h$. 

\begin{figure}[tb]
\centering
\includegraphics[width=\columnwidth]{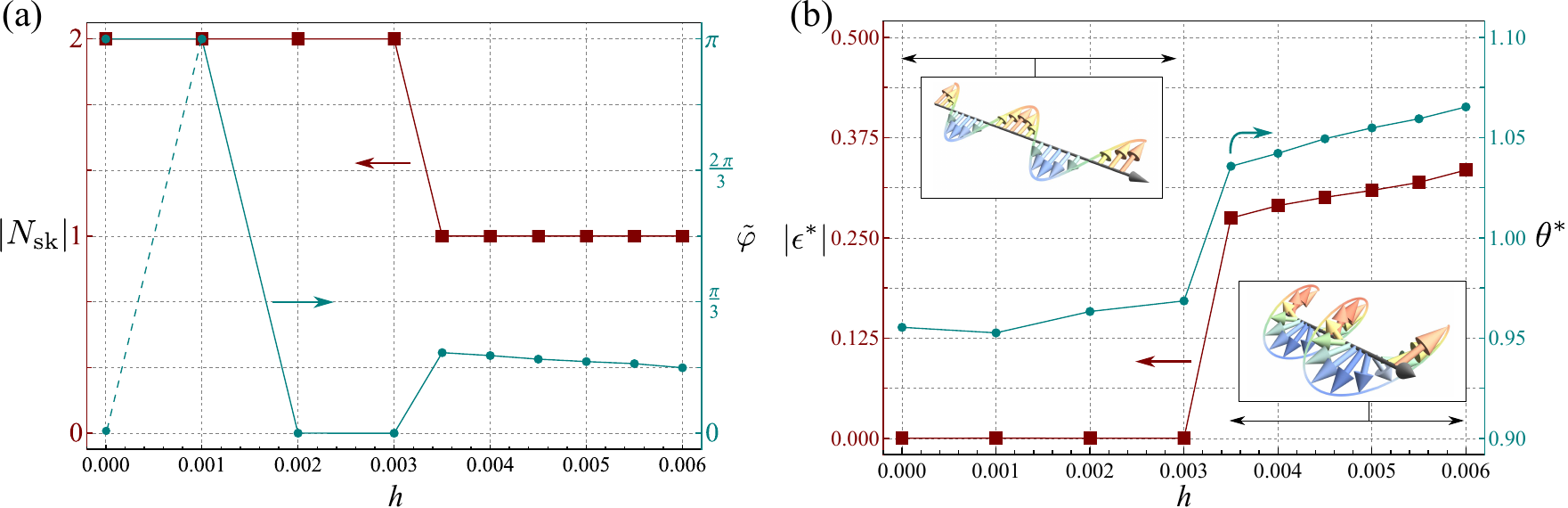}
\caption{\label{fig:2}
(a) Magnetic field dependences of the topological number $|N_{\rm sk}|$ and the sum of phases $\tvp$ in (\ref{eq:phase_sum}). 
(b) Magnetic field dependences of the ellipticity $\epsilon^*$ and the angle $\theta^*$. 
The insets in (b) represent the schematic illustrations of one of three constituent waves.
}
\end{figure}

Figure~\ref{fig:2}(a) shows the skyrmion number per magnetic unit cell, $|N_{\rm sk}|$, calculated for the optimal spin configuration 
$\{{\bf S}_{{\bf r}}^{3Q}\}^*$; there is a phase transition at $h\simeq 0.00325$ accompanied by the topological change from $|N_{\rm sk}|=2$ to $1$. 
The results well reproduce those for $\{{\bf S}_{{\bf r}}^{\rm GS}\}$~\cite{Ozawa2017}. 

The sum of phases $\tvp$ in (\ref{eq:phase_sum}) is also shown in figure~\ref{fig:2}(a). 
For the $|N_{\rm sk}| = 2$ state at $h = 0$, we obtain $\tvp\simeq\pi$, where $\tvp$ is equivalent to $\tvp-\pi$ because of time-reversal symmetry; to
show this explicitly, we plot both points. 
While increasing $h$, we find that $\tvp$ remains $\simeq 0$ in the $|N_{\rm sk}|=2$ region for $h\lesssim 0.00325$, except for $\tvp\simeq\pi$ at $h=0.001$. 
We speculate that this exception is a numerical artifact; the simulation in the previous study presumably failed to find the true ground state with $\tvp\simeq0$ at $h=0.001$, 
as the energy difference between the states with $\tvp\simeq0$ and $\pi$ becomes smaller for lower $h$.  
When entering into the $|N_{\rm sk}|=1$ region for $h\gtrsim 0.00325$, $\tvp$ is suddenly shifted from $0$ to $\sim \frac{\pi}{6}$ 
and gradually decreases while increasing $h$. 
This clearly shows that the phase shift takes place simultaneously with the topological phase transition from $|N_{\rm sk}|=2$ to $1$. 

Figure~\ref{fig:2}(b) shows the ellipticity $\epsilon^*$ and the angle $\theta^*$ in (\ref{eq:3q_fit_ansatz}) as functions of $h$. 
In the $|N_{\rm sk}|=2$ state for $0 \leq h \lesssim 0.00325$, we obtain $\epsilon^*\simeq0$ and $\theta^* \sim \arccos(1/\sqrt{3}) \simeq 0.955$ 
which indicates that the constituent waves are sinusoidal waves as schematically depicted in the upper-left inset 
and the vectors ${\bf e}_{\eta}^{1}(\theta^{*},\Gamma^{*})$ in (\ref{eq:3q_fit_ansatz}) are almost orthogonal to each other. 
On the other hand, both $\epsilon^*$ and $\theta^*$ change suddenly at the topological transition at $h\simeq 0.00325$, and gradually increase while increasing $h$.
The results indicate that the constituent waves in this region are elliptically modulated as exemplified in the lower-right inset 
and the angle between the major axes of the ellipses, ${\bf e}_{\eta}^1(\theta^{*}, \Gamma^{*})$, is larger than $\pi/2$. 
Thus, we find that the topological transition from $|N_{\rm sk}|=2$ to $1$ is accompanied by not only the phase shift but also the changes in the ellipticity and the angle of the propagating planes of the constituent waves.

\section{Concluding remarks \label{sec:4}}
To summarize, we analyzed the real-space spin configurations of two types of triple-$Q$ SkLs previously found 
in the Kondo lattice model, and clarified the phase, the ellipticity, and the angle of the superposed waves. 
We elucidated that these parameters abruptly change at the topological phase transition from $|N_{\rm sk}|=2$ to $1$ 
while increasing the magnetic field. 
We note that our results are qualitatively consistent with those obtained by the recent variational study in \cite{Hayami2021locking}, where the relative phase between the constituent sinusoidal and cosinusoidal waves, which partly corresponds to 
$\tvp$, $\epsilon$, and $\theta$ in our analysis, is shifted at the topological phase transitions induced by the magnetic anisotropy and the magnetic field. 
Our results indicate that the parameters to characterize the constituent waves underlie the change of 
the topological properties in the triple-$Q$ SkLs. 

Our finding provides a first step toward the systematic investigation on other topological phase transitions in terms of 
the detailed parameters in the constituent waves, not only for the SkLs but also for 
other topological spin textures, such as 
the three-dimensional hedgehog lattices~\cite{Binz2006-1,Park2011, Fujishiro2020, Okumura2020}. 
It is desired to establish the mechanism to control the constituent waves, e.g., by the external magnetic field and pressure, in the future studies. 
It is also intriguing to study the relations between the nature of the constituent waves and the electronic, transport, 
and dynamical properties in the topological multiple-$Q$ states.

\ack
We thank R. Ozawa for providing us the numerical data and S. Hayami for fruitful discussions. 
This research was supported by Grant-in-Aid for Scientific Research Grants (Nos. JP18K03447, JP19H05822, JP19H05825, and JP21J20812), JST CREST (No. JP-MJCR18T2), and the Chirality Research Center in Hiroshima University and JSPS Core-to-Core Program, Advanced Research Networks.

\section*{References}
\bibliography{ref}

\end{document}